\documentclass{aa}
\usepackage{natbib}
\usepackage[dvips, pdftex]{graphicx}
\usepackage{epstopdf}
\usepackage{txfonts}
\bibpunct{(}{)}{;}{a}{}{,}

\begin{document}

 \title{An evolutionary study of the pulsating subdwarf B\\eclipsing binary {PG\,1336$-$018} (NY Vir)}

   \author{H.~Hu
          \inst{1,2}
          \and
          G.~Nelemans\inst{1}
          \and
          R.~\O stensen\inst{2}
          \and
          C.~Aerts\inst{1,2}
          \and
         M.~Vu\v{c}kovi\'{c}\inst{2}
          \and
          P.~J.~Groot\inst{1}
          }

   \offprints{H.~Hu}

   \institute{Department of Astrophysics, IMAPP, Radboud University Nijmegen, PO Box 9010, 6500 GL, Nijmegen, the Netherlands\\
              \email{hailihu@astro.ru.nl}
         \and
             Institute of Astronomy, Katholieke Universiteit Leuven, Celestijnenlaan 200D, 3001 Leuven, Belgium\\
             }

   \date{Received --; Accepted--}

 
  \abstract
  {The formation of subdwarf B (sdB) stars is not well understood within the current framework of stellar single and binary evolution.}
   {In this study, we focus on the formation and evolution of the pulsating sdB star in the very short-period eclipsing binary {PG\,1336$-$018}. We aim at refining the formation scenario of this unique system, so that it can be confronted with observations.}
  {We probe the stellar structure of the progenitors of sdB stars in short-period binaries using detailed stellar evolution calculations. Applying this to {PG\,1336$-$018} we reconstruct the common-envelope phase during which the sdB star was formed. The results are interpreted in terms of the standard common-envelope formalism (the $\alpha$-formalism) based on the energy equation, and an alternative description (the $\gamma$-formalism) using the angular momentum equation.}
   {We find that if the common-envelope evolution is described by the $\alpha$-formalism,  the sdB progenitor most likely experienced a helium flash. We then expect the sdB mass to be between $0.39$ and $0.48$ M$_{\odot}$, and the sdB progenitor initial mass to be below $\sim$2 M$_{\odot}$. 
   However, the results for the $\gamma$-formalism are less restrictive, and a broader sdB mass range (0.3 - 0.8 M$_{\odot}$) is possible in this case. Future seismic mass determination will give strong constraints on the formation of {PG\,1336$-$018} and, in particular, on the CE phase.}
   {}


   \maketitle

\section{Introduction}
Subdwarf B (sdB) stars are the dominant population of faint blue objects at high galactic latitudes \citep{green1986}, and are found in both the disk and halo.  They are also ubiquitous in giant elliptical galaxies, where they are believed to be the main source of the ultraviolet excess \citep{brown1997}. In the Hertzsprung-Russell diagram they lie on the blue extension of the Horizontal Branch, and are therefore also known as Extreme Horizontal Branch (EHB) stars. It is generally thought that they are low mass ($0.5$ M$_{\odot}$) core-helium burning stars with extremely thin hydrogen envelopes ($<0.02$ M$_{\odot}$) \citep{heber1986,saffer1994}. Their envelopes are too thin to sustain hydrogen burning, hence they will evolve directly to the white dwarf cooling track after core-helium exhaustion,  without going through the Asymptotic Giant Branch and Planetary Nebulae phases. 

It is not clearly understood how the sdB progenitor manages to loose almost its entire hydrogen-envelope, but nevertheless starts core-helium fusion. Both single star evolution with enhanced mass loss on the Red Giant Branch (RGB) \citep{d'cruz1996}, and binary evolution models \citep{mengel1976} have been proposed as formation channels. Extensive surveys show that a large fraction of sdB stars are in binaries (e.g.~\citealt{allard1994,morales2006}). This motivated \citet{han2002,han2003} to perform a detailed investigation of the main binary evolution channels that can produce an sdB star. They found that an sdB star can be formed after one or two common-envelope (CE) phases producing a short-period binary ($P=$ $\sim$$0.1$ to $\sim$$10$ d) with respectively a main-sequence (MS) or a white dwarf (WD) companion. Only one phase of stable Roche Lobe overflow (RLOF) is predicted to contribute to the sdB population. This channel produces a wider binary  ($P=$ $\sim$$1$ to $\sim$$500$ d) with an MS companion. Single sdB stars are explained by the merger of two helium white dwarfs (WD). The binary population synthesis models for these formation channels \citep{han2003} predict a mass distribution of sdB stars that sharply peaks at the canonical value of 0.46 M$_{\odot}$, but it is much wider ($0.3$ - $0.8$ M$_{\odot}$) than previously assumed.
The wide mass range is due to stars which ignite helium under non-degenerate conditions. These systems had not been explored as sdB progenitors before. 

A fraction of sdB stars show multimode short-period oscillations with amplitudes in the milli-magnitude range. They are observed to have surface gravities ($\log g$) between 5.2 and 6.2 and effective temperatures ($T_{{\rm eff}}$) between 28,000 and 36,000 K. This class of pulsators is known as sdBV or V361 Hya stars. They are also often referred to as EC 14026 stars after the prototype, discovered by \citet{kilkenny1997}. Independently, these oscillations were theoretically predicted to be driven by an opacity mechanism \citep{charpinet}. A seismic study can provide detailed constraints on the sdB interior, most importantly the total mass and the mass of the hydrogen-envelope, which are essential ingredients to tune the sdB formation scenarios. 

An excellent laboratory for a detailed seismic and evolutionary study is the sdB pulsator in the short-period ($2.4$ h) eclipsing binary {PG\,1336$-$018}, also known as NY Vir. The sdB primary was discovered to pulsate by \citet{kilkenny1998}, and has been the target of a Whole Earth Telescope campaign \citep{kilkenny2003}. However, an adequate seismic model has not been determined yet due to the lack of colour information. In future work, we will attempt to achieve this by using high-precision VLT photometry and spectroscopy of this target star. An overview of the data and the orbit solution can be found in \citet{vuckovic}.  Here we  present a progenitor study of {PG\,1336$-$018} from a theoretical point of view. \emph{Assuming} the sdB mass to be the canonical $0.5$ M$_{\odot}$, \citet{kilkenny1998} derived a mass for the companion of $0.15$ M$_{\odot}$ and estimated its class to $\sim$M5V. In view of the wide sdB mass distribution predicted by \citet{han2003}, we drop the assumption on the sdB mass and investigate the range of initial system parameters for binaries that evolve into a {PG\,1336$-$018}-like configuration. 

The current orbital separation, $\sim$0.8 R$_{\odot}$ \citep{vuckovic}, is much smaller than the radius the sdB progenitor had as a red giant. This implies that the system evolved through a common-envelope (CE) and spiral-in phase. At the start of mass transfer, the giant must have achieved a certain minimum core mass, for the core to still ignite helium after loss of the envelope. However, mass transfer must have started before the giant reached the tip of the RGB, because after the tip the giant starts to contract. The range that the giant's core mass and radius can have, has been calculated by \citet{han2002} as a function of the zero-age main-sequence (ZAMS) mass. They showed that the minimum core mass required for helium ignition is typically within 5\% of the core mass at the tip of the RGB, where their definition of the core mass boundary is closely related to the layer of maximum energy production rate \citep{han1994}. 
We perform a similar study here, but with some refined constraints. Most importantly, we take into account that the minimum core mass for helium ignition depends sensitively on the hydrogen envelope that is kept by the star.

While a progenitor study of {PG\,1336$-$018} sheds light on the origin of sdB stars, the future evolution of this system is interesting in the context of cataclysmic variables (CVs). \citet{schreiber2003} calculated that {PG\,1336$-$018} will evolve into a semi-detached configuration within the Hubble-time and thus is representative for progenitors of present-day CVs. However, they mistakenly took the relative radius of the secondary from \citet{kilkenny1998} as the absolute radius. Furthermore, we now have more accurately determined system parameters than \citet{kilkenny1998}. Therefore we reinvestigate the status of {PG\,1336$-$018} as a pre-CV as well and find that the main conclusion by \citet{schreiber2003}  remains true, i.e.~{PG\,1336$-$018} is representative for progenitors of present-day CVs.

The outline of this paper is as follows. Section \ref{evolution} briefly describes the stellar evolution code used in this study, and the procedure we adopted for our calculations. Section \ref{results} presents the results we obtained by probing the stellar structure of progenitors of sdB stars in short-period binaries at the onset of mass transfer. In particular, we present the pre-CE orbital separation of possible progenitors of {PG\,1336$-$018}. This is used in Section \ref{CEphase} to constrain the CE evolution. The results are discussed in Section \ref{discussion} and summarized in Section \ref{conclusions}.

\section{The evolutionary calculations}\label{evolution}
\subsection{The stellar evolution code}\label{code}
We compute the stellar evolution with the numerical computer code originally developed by \citet{eggleton1971, eggleton1972, eggleton1973, faulkner1973} and updated by \citet{han1994} and \citet{pols1995,pols1998}\footnote{A write-up
of the most recent version of this code can be obtained from P.~Eggleton at
ppe@igpp.ucllnl.org.}. The updated version of the code uses an equation of state that includes pressure ionization and Coulomb interaction, opacity tables derived from \citet{rogers1992} and \citet{alexander1994}, nuclear reaction rates from \citet{caughlan1985} and \citet{caughlan1988}, and neutrino loss rates from \citet{itoh1989,itoh1992}. The code uses a self-adaptive, non-Lagrangian mesh. During an iteration it solves implicitly and simultaneously the stellar structure equations, the chemical composition equations and the equations governing the mesh-spacing. Both convective and semi-convective mixing are treated as diffusion processes. \citet{izzard2006} recently developed a graphical user interface, Window To The Stars (WTTS), to Eggleton's code. WTTS significantly simplifies running the code and allows immediate analysis of results.

We use a mixing-length parameter (the ratio of the mixing-length to the local pressure scaleheight) of $\alpha=l/H_p=2.0$. Convective overshooting is included using an overshooting parameter $\delta_{\rm{ov}}=0.12$ which corresponds to an overshooting length of $\sim0.25 H_p$. We use a Reimers' wind mass-loss rate \citep{reimers1975},
\begin{equation}
\dot{M}_{\rm{wind}}=4\times10^{-13}\eta \frac{(R/R_{\odot})(L/L_{\odot})}{(M/M_{\odot})}\textrm{  [M}_{\odot}\textrm{yr}^{-1}],
\end{equation}
 with an efficiency of $\eta=0.4$ \citep{iben1983,carraro1996}. The metallicity is taken to be $Z=0.02$.
 
In \citet{han1994}, the stellar core boundary is related to the layer of maximum energy production rate. This will not be applicable for EHB stars since they have very thin inert hydrogen envelopes. Therefore, we define the core to be the inner region with a hydrogen mass fraction $X<0.10$. This will generally lead to lower values for the core mass, although it will not give a significant difference for degenerate cores \citep{dewi2000,tauris2001}. In principle, it is not important how the core boundary is defined, provided that it not assumed to be the bifurcation point above which all the material is ejected.

\subsection{Procedure}
We have used the Eggleton code to follow the evolution along the RGB of stars with ZAMS masses in the range $1-4$ M$_{\odot}$. We do not study more massive progenitors, because they will result in EHB stars of mass above $0.8$ M$_{\odot}$ \citep{han2002}, which are too hot to become sdB pulsators. We approximate the CE phase by removing the envelope at a rate of $10^{-6} M_{*}$ yr$^{-1}$ where $M_*$ is the mass of the star, while keeping the composition constant. Clearly, this is a crude approximation, as CE evolution involves much higher rates of mass loss causing the star to lose hydrostatic equilibrium. It is, however, expected that the subsequent evolution does not depend on the mass loss history, but it is mainly determined by the amount of hydrogen left. Unfortunately, it is unknown how much envelope is ejected during the CE phase. What we do here is derive an upper limit for this quantity. First of all, we expect the star to contract after CE ejection. However, as sdB spectra are generally dominated by hydrogen lines, we do not remove more envelope than down to $X = 0.10$. Secondly, as we are investigating post-CE sdB stars, the stars cannot have a too large radius after mass-loss, therefore we require $R_{\rm{post-CE}}<10^2$ R$_{\odot}$. This is still a much too large limit for  {PG\,1336$-$018}, but it might apply to other post-CE sdB systems, and for now we want to keep the discussion general. The final condition we impose is that the star must evolve to the EHB with temperatures above $28,000$ K, as is characteristic for pulsating sdB stars.
Which of these three conditions will determine the maximum remaining envelope depends on the situation. To clarify this, we state the four different scenarios with the dominating criterion: 
\begin{itemize}
\item[1)] $M_{\rm{ZAMS}}<2$ M$_{\odot}$, near $M_{\rm{core,min}}$: \\When the core is degenerate, the core boundary is very distinct. In this case, it is reasonable to suppose that nearly all the material above the very compact core is expelled. Also, the large surface gravity at the core boundary will restrict the remaining hydrogen envelope to be $M_{\rm{env}}<$ $\sim$$10^{-3}$ M$_{\odot}$, for exact values see Table \ref{mincore}. We found that thicker envelopes will continue to burn hydrogen and these models have too large radii ($\sim$$150$ R$_{\odot}$) to fit in the narrow orbits of post-CE systems.
\item[2)] $M_{\rm{ZAMS}}<2$ M$_{\odot}$, near RGB tip: \\When the core degeneracy is lifted\footnote{We used approximated post-He-flash models, as the code cannot calculate through the helium flash \citep{pols1998}.  } 
 during the CE ejection, we find that the remaining envelope can have a mass up to $\sim$$10^{-2}$ M$_{\odot}$ (see Table \ref{mincore}), consistent with the evolutionary studies of  \citet{caloi1989} and \citet{dorman1993}. In this case, we determined the maximum envelope which allows the star to reach the EHB with an effective temperature above $28,000$ K.\item[3)] $M_{\rm{ZAMS}}>2$ M$_{\odot}$, near RGB tip: \\When these stars with non-degenerate cores are close to helium ignition, the situation is comparable with case 2. Also here, the maximum envelope mass ($\sim$$10^{-2}$ M$_{\odot}$, see Table \ref{mincore}) follows from the requirement that the sdB star must  reach $T_{\rm{eff}} > 28,000$ K during core helium burning. 
\item[4)] $M_{\rm{ZAMS}}\geq2.5$ M$_{\odot}$, near $M_{\rm{core,min}}$: \\These stars can ignite helium already when they loose their envelopes at the end of the MS. For the end-of-MS  models we have not been able to derive a realistic upper limit for the remaining hydrogen envelope, because they tend to expand even when we remove the entire envelope, i.e.~down to $X=0.10$. For these models, we assumed that the entire envelope was ejected.
\end{itemize}

Using the above conditions for how much envelope should at least be removed, we determined the minimum core mass for helium ignition, and the stellar structure at onset of mass transfer.

\section{Results}\label{results}
\subsection{At the onset of mass transfer}\label{pre-CE}
\subsubsection{The stellar structure}
We are interested in the stellar structure of the sdB progenitors at the onset of mass transfer. In particular the total mass,  the radius, the remnant mass, and the binding energy of the removed envelope are important for constraining the CE evolution. In Table \ref{mincore} we present the relevant stellar parameters corresponding to the minimum core mass for helium ignition and the tip of the RGB.  We also give limits on the total mass and the mass of the hydrogen envelope of the post-CE star, which will be close to the values of the sdB star itself.

Note that for $M_{\rm{ZAMS}}<2$ M$_{\odot}$, the remaining hydrogen envelope can be thicker when mass transfer started at the tip of the RGB. In this case we removed the envelope of post-He-flash models, i.e. we assumed that the core degeneracy was lifted during the CE ejection. For the models at the end of the main-sequence, i.e. $M_{\rm{ZAMS}}>2.5$ M$_{\odot}$ and $M_{\rm{core}}=M_{\rm{core,min}}$, we could not determine the hydrogen envelope reliably (see Section \ref{remenv}). Instead we removed the envelope down to $X=0.10$.

\begin{figure}
   \begin{center}
  \includegraphics[angle=-90, width=9cm]{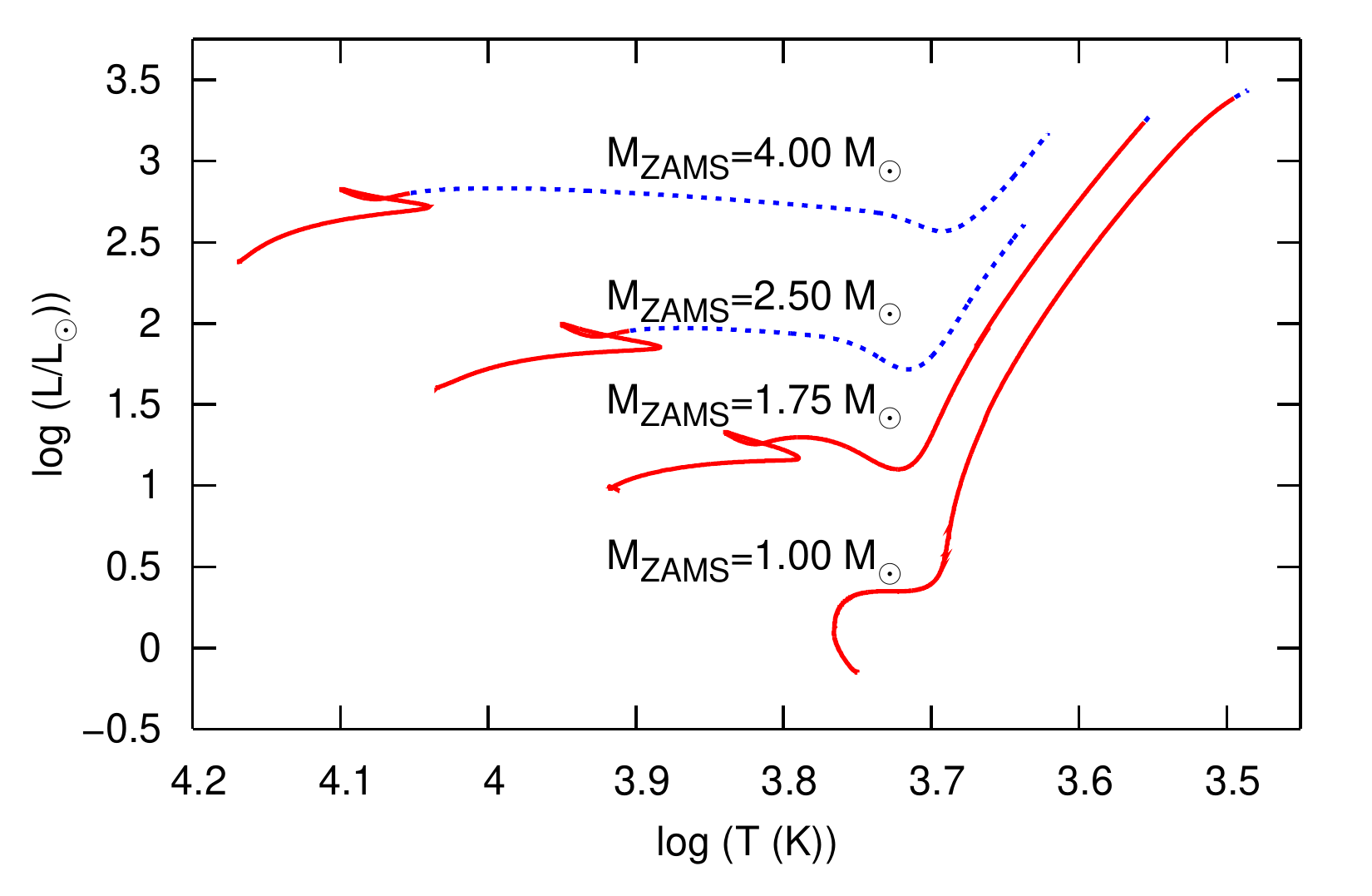}
   \caption{Evolutionary tracks of sdB progenitors until the tip of the RGB in the HR diagram.  At the upper two tracks the red giant ignites helium quiescently, while at the lower two tracks the helium flash occurs. On the dotted part of the track, the core mass is above the minimum required for helium ignition. Thus when the giant looses its envelope in this stage of the evolution, it may become an sdB star. }
              \label{HR}
              \end{center}
    \end{figure}

It is interesting to note that the stellar structure at the onset of mass transfer is quite different depending on whether the giant experienced a helium flash or ignited helium quiescently. For degenerate cores, core contraction is inhibited by the degeneracy pressure, thus mass transfer can only have started very near the tip of the RGB, for the core to still ignite helium. More massive stars ($M_{\rm{ZAMS}}>2$ M$_{\odot}$) achieve temperatures high enough to avoid core degeneracy. They do not have to be close to the tip of the RGB at the onset of mass transfer to still ignite helium, as long as their core mass exceeds the absolute minimum for helium ignition ($\sim$$0.3$ M$_{\odot}$). \citet{han2002} found that stars with $M_{\rm{ZAMS}}\geq2.5$ M$_{\odot}$ will burn helium even when the envelopes are lost when passing through the Hertzsprung gap.\footnote{Stars with $2<M_{\rm{ZAMS}}<2.5$ M$_{\odot}$ leave the main-sequence with core masses $\sim$0.25 M$_{\odot}$ and  do not achieve their minimum core mass until they are quite near the RGB tip.} This can be clearly seen in the HR diagram shown in Fig.~\ref{HR}.

\begin{table*}
\begin{minipage}{\textwidth}
\caption{The stellar structure before and after mass transfer.\protect\footnote{For each ZAMS mass, the first line gives the stellar structure corresponding to the minimum core mass for helium ignition. The second line corresponds to the tip of the RGB. The columns are respectively: $M_{\rm{ZAMS}}$ = zero-age main-sequence mass; $M_*$ = total mass of giant; $M_{\rm{core}}$ = helium core mass of giant, $R_*$ = radius of giant; $\lambda_{\rm{gr}}$ and  $\lambda_{\rm{tot}}$ are dimensionless parameters indicating respectively the gravitational binding energy and the total (including thermal) binding energy of the ejected envelope (see Section \ref{ALPHA}); $I$ = the moment of inertia; $M_{\rm{env}}$ = hydrogen envelope left after CE ejection; $M_{\rm{remnant}}$ = remnant mass after CE ejection. }} \label{mincore}
\begin{center}
\begin{tabular}{cl|ccrccc|ll}
\vspace{-0.25cm}\\
\hline
\vspace{-0.25cm}\\
&&\multicolumn{6}{c}{pre-CE}&\multicolumn{2}{c}{post-CE}\\
\hline
\vspace{-0.25cm}\\
$\frac{M_{\rm{ZAMS}}}{M_{\odot}}$&&$\frac{M_{*}}{M_{\odot}}$&$\frac{M_{\rm{core}}}{M_{\odot}}$&$\frac{R_{*}}{R_{\odot}}$&$\lambda_{\rm{gr}}$&$\lambda_{\rm{tot}}$&$\frac{I}{M_*R_*^2}$&$\frac{M_{\rm{env}}\times 10^{-3}}{M_{\odot}}$&$\frac{M_{\rm{remnant}}}{M_{\odot}}$\\
\vspace{-0.25cm}\\
\hline
\vspace{-0.25cm}\\
\multicolumn{10}{l}{helium flash}\\
\vspace{-0.25cm}\\
1.00&min&0.821&0.460&168&0.55&4.61& 0.052&$\leq1.3$&0.46 \\
 &tip&0.791&0.472&185&0.53&4.28&0.051&$\leq10$&0.47 - 0.48\\ 
1.50&min&1.401&0.454&136&0.64&3.74&0.084&$\leq1.4$&0.45 \\ 
 &tip&1.395&0.466&148&0.57&3.36&0.082&$\leq9$&0.47 - 0.48\\ 
1.75&min&1.695&0.434&107&0.69&2.92&0.097&$\leq1.7$&0.43 \\  
&tip&1.687&0.446&118&0.62&2.70&0.095&$\leq7$&0.45 - 0.46\\   
1.95&min&1.926&0.394&65&0.73&2.18&0.113&$\leq2.7$&0.39 \\  
&tip&1.926&0.394&65&0.72&2.14&0.113&$\leq6$&0.39 - 0.40\\  
\vspace{-0.25cm}\\
\multicolumn{9}{l}{non-degenerate helium ignition}\\
\vspace{-0.25cm}\\
2.05&min&2.045&0.317&26&0.83&1.90&0.135&$\leq14$&0.32 - 0.33\\ 
 &tip&2.043&0.320&26&0.83&1.92&0.135&$\leq13$&0.32 - 0.33\\ 
2.50&min&2.494&0.322&5&0.25&0.47&0.022&-&-\\  
&tip&2.493&0.372&36&0.85&2.02&0.134&$\leq34$&0.37 - 0.41\\  
3.00&min&2.993&0.411&6&0.24&0.46&0.021&-&- \\  
 &tip&2.993&0.444&45&0.84&2.00&0.132&$\leq61$&0.44 - 0.50\\  
4.00&min&3.991&0.596&8&0.26&0.48&0.022&-&-\\   
 &tip&3.990&0.623&74&0.78&1.94&0.125&$\leq116$&0.62 - 0.74\\   
\hline
\end{tabular}
\end{center}
\end{minipage}
\end{table*}

\subsubsection{The orbital separation}\label{orb}
It is generally assumed that mass transfer starts as soon as the giant fills its Roche lobe, i.e.~when its radius equals the Roche radius. The Roche radius is approximated by
\begin{equation}\label{rocheradius}
R_L=\frac{0.49q^{2/3}a}{0.6q^{2/3}+\ln(1+q^{1/3})},
\end{equation}
where $q=M_1/M_2$ is the mass ratio and $a$ the orbital separation \citep{eggleton1983}. However, if the binary system is tidally unstable \citep{counselman1973}, the Roche geometry is not applicable. It is expected that the time-scale of tidal evolution is much shorter than the time-scale of nuclear expansion, i.e. when a tidal instability sets in, the stars spiral inwards before the stellar structure can change significantly. Therefore, when a tidal instability sets in before the giant fills its Roche lobe, this can also cause a CE. The binary becomes tidally unstable when the spin angular momentum of the stars exceeds one third of the orbital angular momentum $h$:
\begin{equation}\label{tidal}
(I_1+I_2)\omega>\frac{1}{3}h\qquad\textrm{with}\qquad h=\sqrt{\frac{GaM_{\rm{giant}}^2M_2^2}{M_{\rm{giant}}+M_2}},
\end{equation}
where $I_1$ and $I_2$ are the moments of inertia of the stars and $\omega$ is the angular velocity  \citep{hut1980}. Since the giant is much larger and more massive than the MS companion, we have $I_1\gg I_2$. The moment of inertia of the giant, $I_1$, follows from detailed evolutionary calculations, and can be found in Table \ref{mincore}. 

\begin{figure}
   \begin{center}
  \includegraphics[angle=-90, width=9cm]{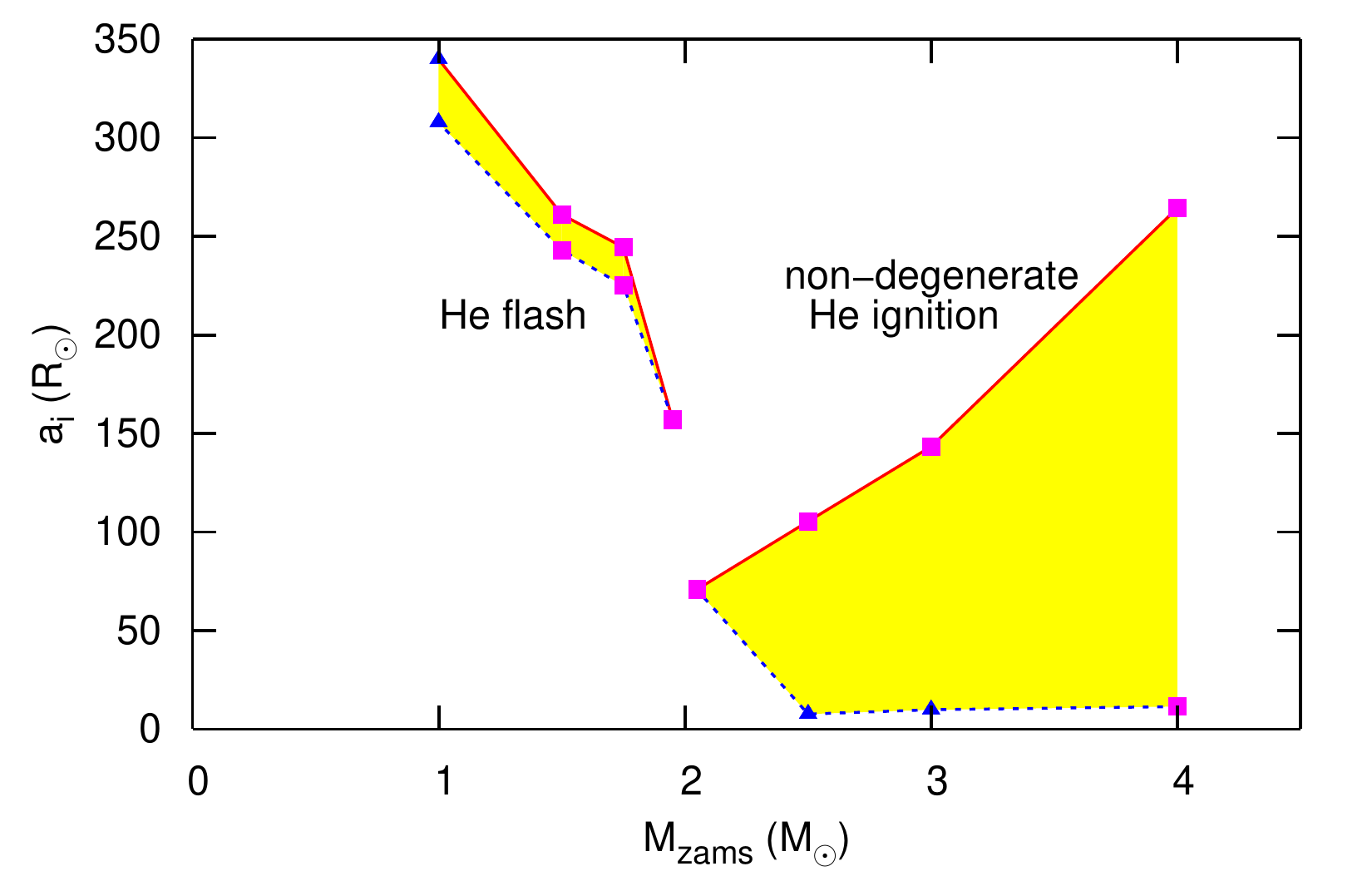}
   \caption{The pre-CE orbital separation $a_i$ as a function of the primary ZAMS mass for $M_2=0.12$ M$_{\odot}$. At the dotted line, mass transfer started when the helium core mass reached the minimum required for helium ignition. At the solid line, mass transfer started at the tip of the RGB. Thus the shaded area gives the possible values of $a_i$ for {PG\,1336$-$018}. At the triangles, the CE is formed by dynamically unstable RLOF. At the squares, the CE is caused by a spiral-in due to a tidal instability.}
              \label{orb_per2}
              \end{center}
    \end{figure}

Using the allowed range for the radius and mass of the primary at the onset of mass transfer (Table \ref{mincore}), we can calculate the pre-CE orbital separation $a_i$ for a given secondary mass. If the binary is tidally stable when it fills its Roche lobe, Eq.~(\ref{rocheradius}) can be used to obtain $a_i$. Otherwise, the CE is formed by a tidal instability and we use
\begin{equation}
I_1\omega=\frac{1}{3}h.
\end{equation}
In Fig.~\ref{orb_per2}, the range of $a_i$ is given as a function of the primary ZAMS mass. We have chosen a secondary mass of 0.12 M$_{\odot}$ here, which is a best-fit value for {PG\,1336$-$018}, observationally derived  by \citet{vuckovic}. The effect of a different secondary mass is investigated in section \ref{secmass}. 

\subsection{Common-envelope ejection}\label{CEphase}
\subsubsection{The energy equation: the $\alpha$-formalism}\label{ALPHA}
In the original spiral-in picture \citep{paczynski1976}, the companion experiences drag forces as it moves into the envelope of the giant and spirals inwards. This idea is applicable when a tidal instability causes the CE or sets in soon after the CE is formed, i.e.~when the mass ratio is high. In this case, the CE is not in co-rotation and the envelope is heated and unbound by friction. The orbital energy $E_{\rm{orb}}$ released in the spiral-in process, is used to eject the envelope with an efficiency $\alpha$,
\begin{equation}\label{Eorb}
\alpha (E_{\rm{orb,f}}-E_{\rm{orb,i}})=E_{\rm{env}},
\end{equation}
where $E_{\rm{env}}$ is the binding energy of the ejected envelope, and the subscripts $i$ and $f$ denote values before and after the CE phase. In principle, one expects $0<\alpha\leq1$. However, in order to explain observed binaries one often finds that $\alpha$ exceeds unity. This might indicate that other energy sources can contribute to the ejection of the envelope, e.g.~the luminosity of the giant. Still, a very high value for $\alpha$ is not anticipated, since it would be physically difficult to explain where such a large amount of energy could come from. The poorly understood physics of the CE phase does not allow us to set a hard limit on $\alpha$, but here we assume $0<\alpha<$ $\sim$5 to be realistic.

It is reasonable to suppose that  the secondary did not accrete any matter since the mass transfer time-scale is short. Expression (\ref{Eorb}) can then be written as
\begin{equation}\label{alphalambda}
\alpha\Big(\frac{-GM_{\rm{remnant}}M_{2}}{2a_{\rm{f}}}+\frac{GM_{\rm{giant}}M_{2}}{2a_{\rm{i}}}\Big)=-\frac{GM_{\rm{giant}}M_{\rm{env}}}{\lambda R_{\rm{giant}}},
\end{equation}
where we have expressed $E_{\rm{env}}$ in terms of the structural parameter $\lambda$  \citep{webbink1984}. It is straightforward to calculate the combined parameter $\alpha\lambda$ from Eq.~(\ref{alphalambda}). To isolate $\alpha$ one usually takes $\lambda=0.5$ \citep{dekool1990}, but an exact calculation should take into account that $\lambda$ depends on the stellar structure. 

\begin{figure}
   \begin{center}
  \includegraphics[angle=-90, width=9cm]{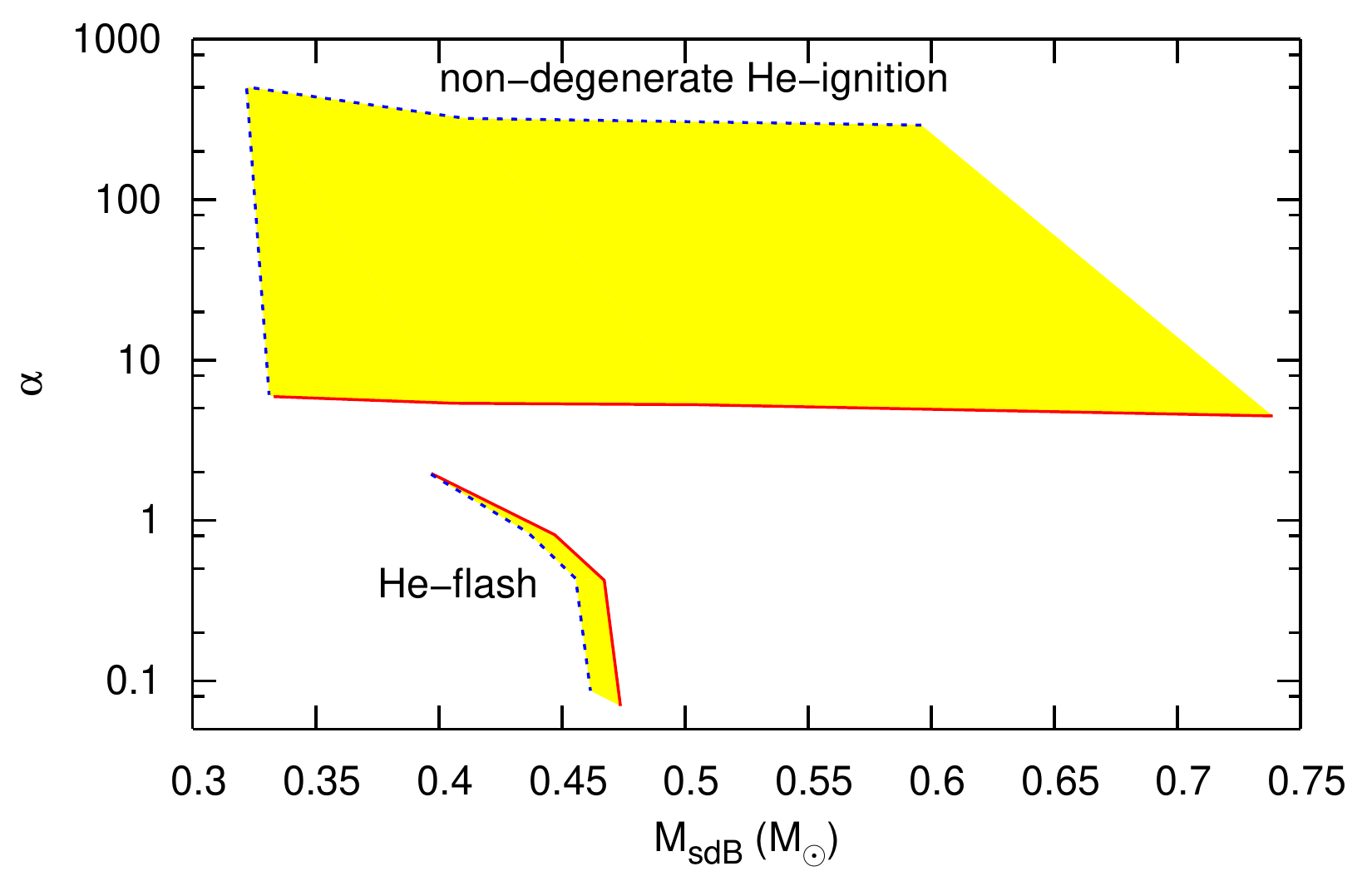}
   \caption{The CE parameter $\alpha$ as a function of the sdB mass for $M_{2}=0.12$ M$_{\odot}$. At the dotted line, mass transfer started when the helium core reached the minimum required for helium ignition. At the solid line, mass transfer started at the tip of the RGB. The shaded region in between indicated the possible $\alpha$-values for  {PG\,1336$-$018}. We assumed here that the binding energy of the envelope is determined by the gravitational energy, i.e.~ $\lambda=\lambda_{\rm{gr}}$.}
              \label{alphamintip}
              \end{center}
    \end{figure}

\begin{figure}
   \begin{center}
  \includegraphics[angle=-90, width=9cm]{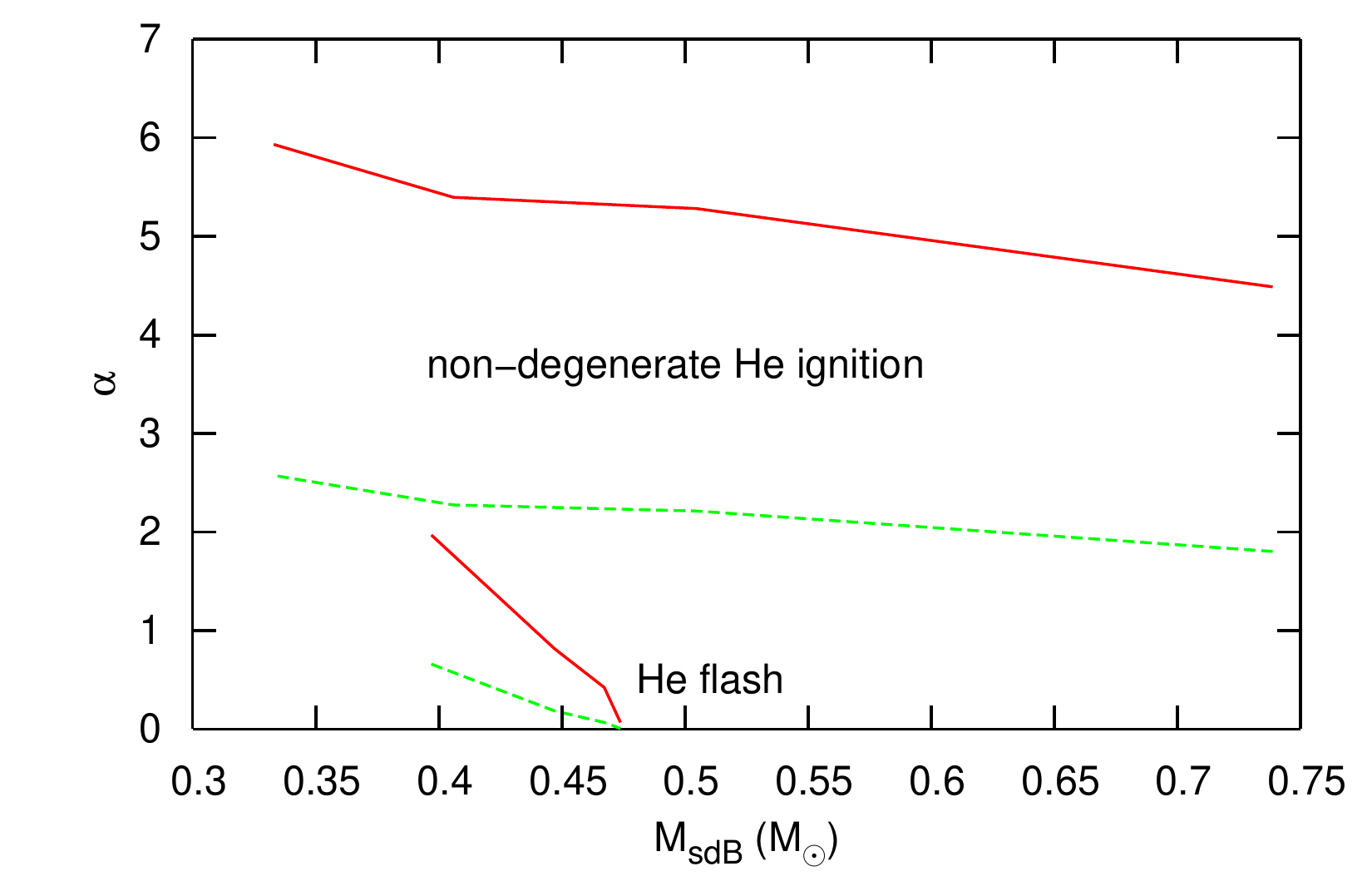}
   \caption{The CE parameter $\alpha$ as a function of the sdB mass for $M_{2}=0.12$ M$_{\odot}$. At the upper solid line, we have $\lambda=\lambda_{\rm{gr}}$ and at the lower dashed line, we have $\lambda=\lambda_{\rm{tot}}$. We assumed that mass transfer started when the red giant reached the tip of the RGB.}
              \label{alphatip}
              \end{center}
    \end{figure}

The total binding energy consists of the gravitational binding energy and the internal thermodynamic energy $U$,
\begin{equation}
E_{\rm{bind}}=\int_{M_{\rm{remnant}}}^{M_{\rm{giant}}}\Big(-\frac{GM}{r}+U\Big)dm.
\end{equation}
Taking the total binding energy of the envelope to calculate $\lambda$ implies that the entire internal energy is used efficiently in the ejection process. However, it is uncertain how much of the internal energy contributes to the ejection of the envelope. This uncertainty is expressed in a parameter  $\alpha_{\rm{th}}$, introduced by \citet{han1994},
\begin{equation}\label{bind}
E_{\rm{env}}=\int_{M_{\rm{remnant}}}^{M_{\rm{giant}}}\Big(-\frac{GM}{r}\Big)dm+\alpha_{\rm{th}}\int_{M_{\rm{remnant}}}^{M_{\rm{giant}}}Udm.
\end{equation}
Expression (\ref{bind}) can be regarded as the effective binding energy of the envelope and is used to derive $\lambda$. We calculated $\lambda$ for $\alpha_{\rm{th}}=1$ and $\alpha_{\rm{th}}=0$, representing respectively the total binding energy ($\lambda_{\rm{tot}}$) and the gravitational binding energy  ($\lambda_{\rm{gr}}$), see Table \ref{mincore}. 
In Fig.~\ref{alphamintip}, $\alpha$ is plotted as a function of the sdB mass for $\lambda=\lambda_{\rm{gr}}$. In principle, we do not expect the ionization energy to contribute to the ejection of the CE, as \citet{harpaz1998} argued that after recombination the opacity drops sharply, hence the released energy flows outward without unbinding the material. In the case that the internal energy of the envelope does contribute to the ejection process (i.e.~$\alpha_{\rm{th}}=0$ and $\lambda=\lambda_{\rm{gr}}$), this will result in a lower ($\sim$factor 2) value of $\alpha$, as is expected from the virial theorem. This can be seen in Fig.~\ref{alphatip}.

It is shown that, in general, non-degenerate helium ignition requires unphysically high $\alpha$-values ($\alpha>5$). However, if the internal energy of the envelope is efficiently used to unbind the envelope, we find $\alpha>2$. Using the initial mass function  $\Phi(M)\propto M^{-2.7}$ \citep{kroupa1993} and the distribution in orbital separation $\Gamma(a)\propto a^{-1}$ \citep{kraicheva1978}, we found that the number of systems that experienced a helium flash is comparable to the number of systems that ignited helium non-degenerately with $\alpha<5$. Thus, if the internal energy can unbind the envelope, we are less confident in excluding the non-degenerate scenario.

    
\subsubsection{The angular momentum equation: the $\gamma$-formalism}\label{GAMMA}
Nelemans et al.~(2000, 2005) found that the first phase of mass transfer of observed double white dwarfs cannot be described by the standard $\alpha$ formalism, nor by stable RLOF. They argued that, for binaries with mass ratio close to unity,  the common envelope is formed by a runaway mass transfer rather than a decay of the orbit. In this case, the angular momentum of the orbit is so large that the common envelope is brought into co-rotation. Consequently, there are no drag forces that can convert orbital energy into heat and kinetic energy. \citet{nelemans2000} described this scenario in terms of the angular momentum balance, the $\gamma$-formalism. 
The assertion is that the specific orbital angular momentum carried away by the envelope is $\gamma$ times the initial specific orbital angular momentum,
\begin{equation}
\gamma \frac{J_{\rm{i}}}{M_{\rm{giant}}+M_{2}}=\frac{J_{\rm{i}}-J_{\rm{f}}}{M_{\rm{env}}}.
\end{equation}
Although the $\gamma$-formalism was originally developed for double white dwarfs, \citet{nelemans2005} showed that most observed sdB binaries can also be explained by $\gamma\sim1.5$. The physical motivation for this empirical description might be super-Eddington mass transfer \citep{beer2007}. Originally this idea was put forward to explain systems in which a main-sequence star transfers mass to a neutron star or black hole \citep{king1999}. \citet{beer2007} proposed that this physical picture is also applicable to systems where a red giant overflows onto a main-sequence star. Their treatment gives an upper limit for $\gamma$, since the specific angular momentum carried away by the envelope cannot be higher than the specific angular momentum of the secondary,
\begin{equation}
\gamma_{\rm{max}}=\frac{M_{\rm{giant}}+M_2}{M_{\rm{env}}}-\frac{(M_{\rm{giant}}+M_2)^2}{M_{\rm{env}}(M_{\rm{core}}+M_2)}\exp\Big(-\frac{M_{\rm{env}}}{M_2}\Big).
\end{equation}

\begin{figure*}
  \begin{center}
  \includegraphics[angle=-90, width=18cm]{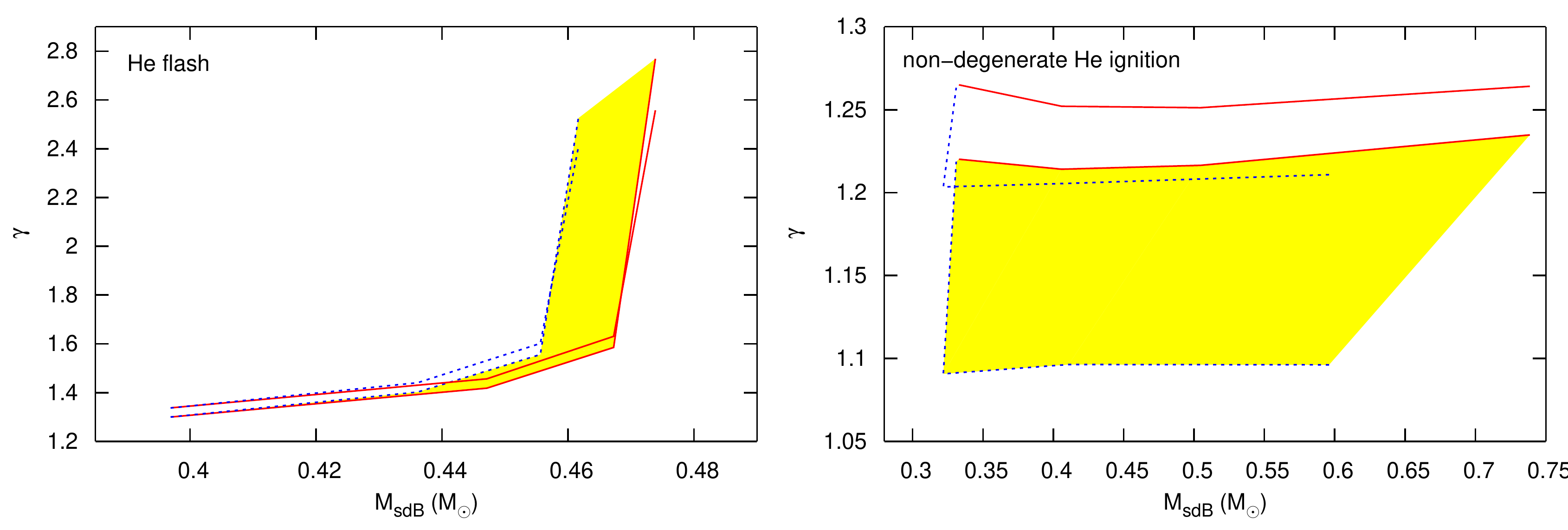}
   \caption{The parameter $\gamma$ as a function of the sdB mass for $M_2=0.12$ M$_{\odot}$. For visibility we have plotted the $\gamma$-values for stars that ignited helium (non-)degenerately in the (right) left panel. At the lower dotted line, mass transfer started when the helium core mass reached the minimum required for helium ignition. At the lower solid line, mass transfer started at the tip of the RGB. The shaded region in between indicated the possible $\gamma$-values for  {PG\,1336$-$018}. The upper dotted line corresponds to $\gamma_{\rm{max}}$ at the minimum core mass, and the upper solid line to $\gamma_{\rm{max}}$ at the RGB tip.}
              \label{gamma}
    \end{center}
    \end{figure*}

We have calculated the possible $\gamma$-values for {PG\,1336$-$018} and plotted them in Fig.~\ref{gamma} as a function of the sdB mass. We also plotted $\gamma_{\rm{max}}$ corresponding to the case that the envelope is ejected with the specific angular momentum of the secondary. 

The common value $\gamma\sim1.5$ is found in the helium flash region, but when the sdB mass approaches the canonical value of $0.47$ M$_{\odot}$, the $\gamma$-value increases steeply to $>\gamma_{\rm{max}}$. This is because the corresponding progenitor mass ($M_{\rm{ZAMS}}=1$ M$_{\odot}$) and the expelled envelope mass are low, thus the angular momentum carried away per unit mass is high. If the progenitor mass of  {PG\,1336$-$018} is higher than $1$ M$_{\odot}$, the $\gamma$-values agree with the idea that the envelope is ejected with almost the specific angular momentum of the secondary. For non-degenerate helium ignition we find $1.1<\gamma<1.2$, and we cannot rule out this possibility as an explanation for {PG\,1336$-$018}.
\section{Discussion}\label{discussion}

\subsection{The influence of the secondary mass}\label{secmass}  
\begin{figure*}
  \begin{center}
  \includegraphics[angle=-90, width=18cm]{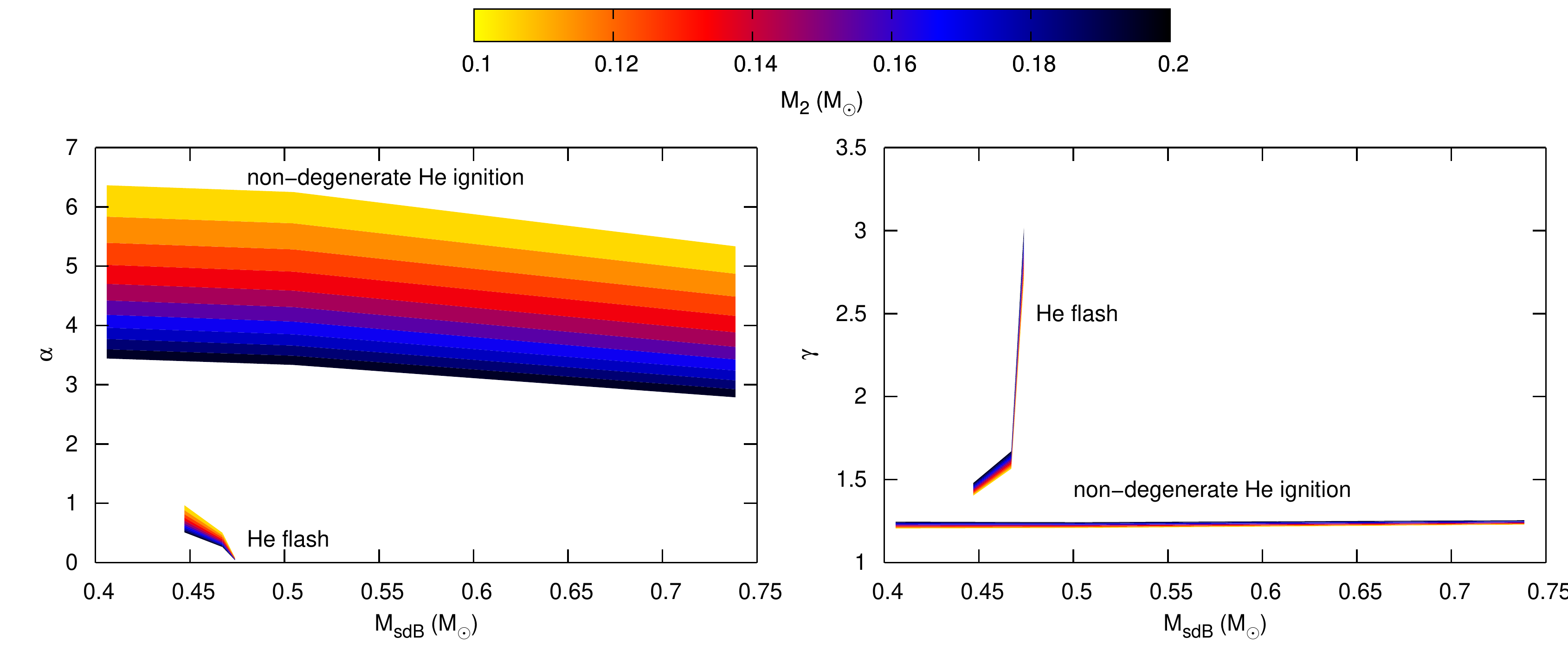}
   \caption{The parameters $\alpha$ and $\gamma$ as a function of the sdB mass, plotted respectively in the left and right panel. The colour gradient gives the secondary mass. We assumed that mass transfer started at the tip of the RGB, and $\lambda=\lambda_{\rm{gr}}$.}
              \label{ag_sec}
    \end{center}
    \end{figure*}

Assuming Newtonian mechanics, the radius of the secondary is given by
\begin{equation}\label{R2}
R_2=\frac{R_1r_2}{r_1}=\sqrt{\frac{GM_{\rm{sdB}}}{g}}\frac{r_2}{r_1}.
\end{equation}
The best-fit orbital solutions of \citet{vuckovic} give averaged relative radii $r_1=0.19$, $r_2=0.21$, and $\log g=5.77$. Substituting these values and the sdB mass range $M_{\rm{sdB}}\sim0.3$ - $0.8$ M$_{\odot}$ in Eq.~(\ref{R2}) gives $R_2\sim0.13-0.21$ R$_{\odot}$. Theoretical models of low-mass main-sequence stars by \citet{baraffe1998} predict the mass range $M_2\sim 0.10-0.20$ M$_{\odot}$ for these radii. In Fig~\ref{ag_sec}, the influence of the secondary mass on the $\alpha$- and $\gamma$-parameter is depicted. We see that $\alpha$ is lower when the secondary is more massive. However, the best-fit orbit solutions of \citet{vuckovic} all have low masses for the secondary ($M_2=0.11-0.13$ M$_{\odot}$). The secondary mass has negligible influence on the $\gamma$-parameter.    

\subsection{Similar systems}
Up to now 39 sdBV stars are known, of which 16 are known
binaries. 13 have spectroscopic F-G companions, two have invisible
WD companions, detectable only by radial velocity variations of
the sdB star, and only {PG\,1336$-$018} has been found to have
an M-dwarf companion. Additionally, one star has been found to
have a long period Jupiter mass planet \citep[{HS\,2201+2610}][]{Silvotti07}
detected only by the variation of the main pulsation period.
Of the remaining 22 stars only four have been carefully checked
for RV variations, the rest remains unexplored.

Since {PG\,1336$-$018} is the primary of an eclipsing system,
it is an exceptionally promising candidate for a study of the sdB
binary formation channel.
A few more non-pulsating sdB binary systems with an M dwarf
companion have been detected (see Table \ref{sdB}), either through eclipses or strong
reflection effects.  The eclipsing binaries {HW\,Vir}
\citep{menzies1986} and {HS\,0705+6700}
\citep{drechsel2001} show striking similarities with {PG\,1336$-$018}. 
The fourth known eclipsing system, {HS\,2231+2441}, while similar
with respect to the short orbital period and other light-curve
properties, turns out to have a very low mass M dwarf companion, just at the substellar limit \citep{ostensen07a} \footnote{We note that this system might also be a non-He-burning post-RGB star with a substellar companion \citep{ostensen07b}, but for our discussion we will assume it is an sdB system.}. A fifth very similar sdB+dM system was also recently detected \citep{Polubek07} in the OGLE-II photometry of the galactic bulge \citep{Wozniak02}, but no spectroscopic data has yet been obtained that can constrain the physical parameters of this system.

The very narrow distribution in orbital periods and component masses
suggests that there is a common mechanism for producing such systems
and that the CE phase of these systems must have been very similar.
To illustrate this, we have calculated $\alpha$ and $\gamma$
for the systems {HW Vir}, {HS\,0705+6700} and {HS\,2231+2441}. For simplicity, we assumed here that the entire envelope was lost when the giant was at the tip of the RGB, and that the internal energy of the envelope cannot be used to unbind the envelope (i.e.~$\alpha_{\rm{th}}=0$ and $\lambda=\lambda_{\rm{gr}}$). 

We can see in Table \ref{sdB} that the CE parameters are indeed quite
similar for these systems. The exception is {HS\,2231+2441}, due to its very low mass companion. Note again that the canonical sdB mass ($0.47-0.48$ M$_{\odot}$) corresponds to high $\gamma$-values.

\begin{figure}[t]
\centering
\includegraphics[width=\hsize]{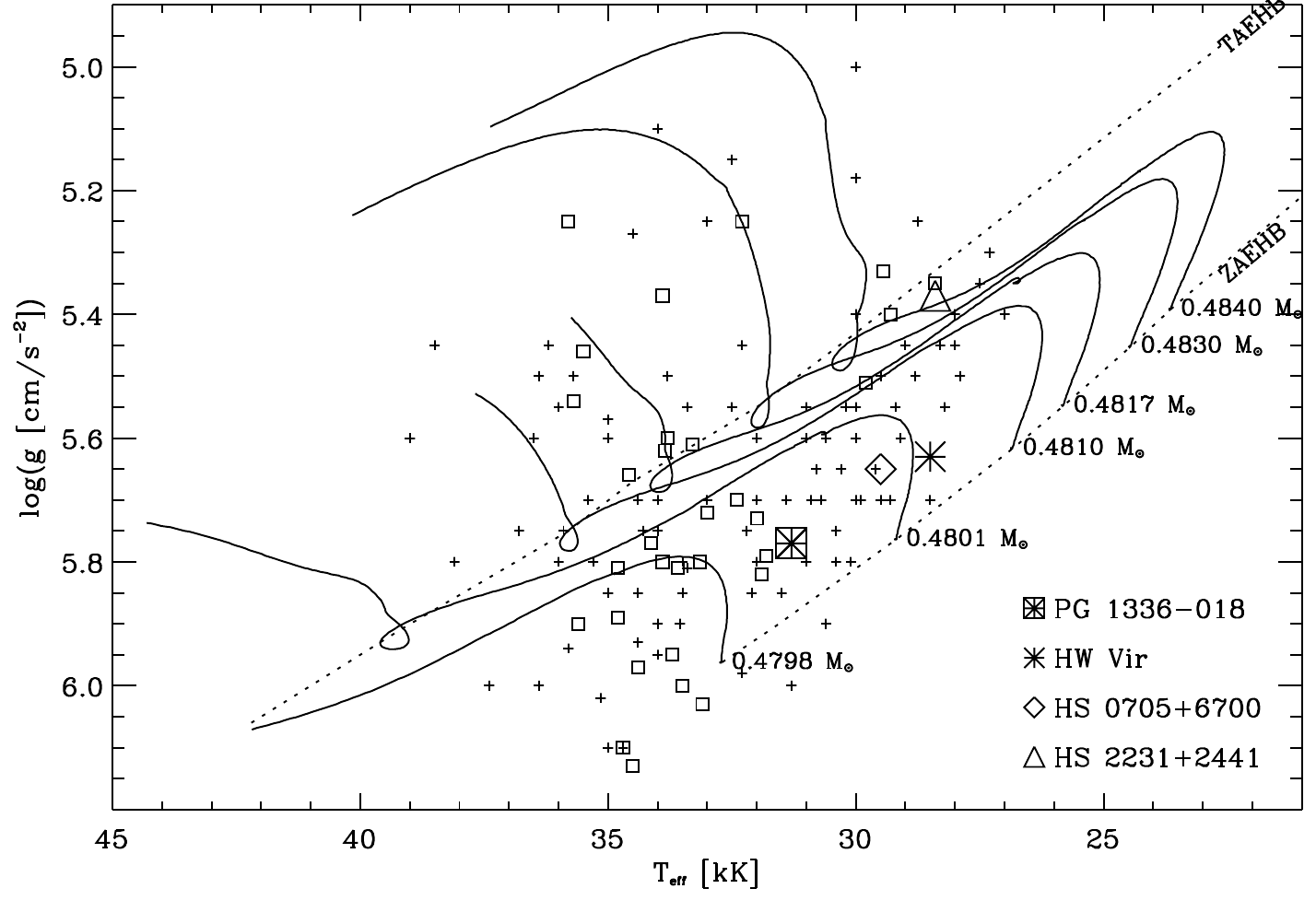}
\caption{The location of {PG\,1336$-$018} in the $T_{\rm eff}$--log $g$ plane plotted together with the other known eclipsing systems, non-eclipsing reflection binaries and some of the pulsating and non-pulsating sdB stars. The squares show known pulsators and the + symbols non-pulsators. Also shown are evolutionary tracks for a sample of EHB stars \citep{kawaler2005}.}
\label{fig:tgplot}
\end{figure}

Fig.~\ref{fig:tgplot} shows the position of {PG\,1336$-$018} in a $T_{{\rm eff}}-\log g$ diagram, together with the
other three systems of HW Vir type for which a spectroscopic temperature
and gravity have been derived, among other sdB stars. Evolutionary tracks from \citet{kawaler2005} are also shown. We see that if {HS\,2231+2441} as an sdB system, it is quite evolved on the EHB, while the other three systems are still in their first half of the EHB evolution.


\begin{table*}
\begin{minipage}{\textwidth}
\caption{System parameters of sdB+dM binaries similar to {PG\,1336$-$018}. \protect\footnote{The $\alpha$- and $\gamma$-values given here correspond to the case that the entire hydrogen envelope was ejected at the tip of the RGB. Whenever the sdB mass allows it, we assumed that helium was ignited degenerately in a flash, indicated by normal font style. The CE parameters in boldface correspond to non-degenerate helium ignition. If $I_1\omega/h<1/3$, the CE is assumed to be formed by dynamically unstable RLOF. If $I_1\omega/h>1/3$, the CE is assumed to be caused by a tidal instability.} }
\label{sdB}
\begin{center}
\begin{tabular}{lllll|lll}
\hline
\vspace{-0.25cm}\\
& Ref.&$P$ (d)& $M_{\rm{sdB}}$ (M$_{\odot}$) & $M_2$ (M$_{\odot}$) &$I_1\omega/h$&$\alpha$&$\gamma$\\
\vspace{-0.25cm}\\
\hline
\vspace{-0.25cm}\\
{HW\,Vir} & \citet{wood1999}& 0.117 & 0.48 & 0.14 &0.10&0.34&2.81 \\
{PG\,1336$-$018} &\citet{vuckovic}&0.101&0.39&0.11 &0.80&3.09&1.29\\
&&&0.47&0.12&0.11&0.36&2.76\\
&&&0.53&0.13&0.17&\textbf{14.4} &\textbf{1.18}\\
{HS\,0705+6700}  & \citet{drechsel2001}& 0.096 & 0.48& 0.13&0.10&0.32&2.79 \\
{HS\,2231+2441} & \citet{ostensen07a}&0.111 & 0.47& 0.075 &0.20&0.59&2.61\\
\vspace{-0.25cm}\\
\hline
\end{tabular}
\end{center}
\end{minipage}
\end{table*}


\subsection{The future evolution of {PG\,1336$-$018}}
Following \citet{schreiber2003} but using updated system parameters, we determine the expected future evolution of {PG\,1336$-$018}. The orbital period will decrease as the system loses angular momentum,
\begin{equation}\label{J}
\frac{\dot{J}}{J}=\frac{\dot{P}}{3P}.
\end{equation}
For $M_2<0.3$ M$_{\odot}$, the secondary is fully convective and angular momentum loss via magnetic braking is negligible \citep{verbunt1981}. Thus we only need to consider angular momentum loss due to gravitational radiation that follows from Einstein's quadrupole formula, 
\begin{equation}\label{gr}
\dot{J}=-\frac{32G^{7/3}}{5c^5}M_1^2M_2^2(M_1+M_2)^{-2/3}\Big(\frac{2\pi}{P}\Big)^{7/3}.
\end{equation}
Eventually the secondary will fill its Roche lobe, initiating a second RLOF phase. As the secondary is essentially unevolved, we assume that its expansion due to its nuclear evolution is negligible. Using Equation (\ref{rocheradius}), Kepler's third law and the system parameters obtained by \citet{vuckovic}\footnote{They found three statistically equivalent solutions, here we use their Model II. The results will not differ significantly for the other two models.}, $M_1=0.466\textrm{ M}_{\odot}$, $M_2=0.122\textrm{ M}_{\odot}$, $R_1=0.15\textrm{ R}_{\odot}$, 
$R_2=0.16\textrm{ R}_{\odot}$, 
the orbital period of the semi-detached system can be calculated: $P_{\rm{sd}}=0.069$ d. The time from now until the secondary fills its Roche lobe follows from equations (\ref{J}) and (\ref{gr}),
\begin{equation}
t_{\rm{sd}}=\frac{5c^5}{256G^{5/3}(2\pi)^{8/3}}\frac{(M_1+M_2)^{1/3}}{M_1M_2}(P_{\rm{now}}^{8/3}-P_{\rm{sd}}^{8/3}).
\end{equation}
We find $t_{\rm{sd}}=9.9\times 10^8$ yr. Since the time-scale on which the sdB star evolves into a white dwarf is a few $10^8$ yr, {PG\,1336$-$018} will evolve into a short-period cataclysmic variable (CV). The present age of {PG\,1336$-$018} is maximally ${12}$ Gyr. This is for $M_{\rm{ZAMS}}=1$ M$_{\odot}$ and the age decreases steeply with increasing mass. Thus the maximum total time until the system becomes semi-detached is less than the Hubble time. We conclude that {PG\,1336$-$018} is representative for progenitors of present-day CVs. Not surprisingly, we find that the similar systems given in Table \ref{sdB} are also pre-CV candidates.

\subsection{Determining the minimum core mass and the remaining hydrogen envelope}\label{remenv}
We cannot model the detailed physics of CE evolution as this is a hydrodynamical event. As an approximation, we used a low enough mass loss rate to ensure numerical stability, and prevent composition changes due to nuclear burning. However, in our models the composition profile does change slightly during the removal of the envelope. We are not sure if this is merely a numerical effect, or that some convective mixing might occur during CE ejection. In any case, we checked that resetting the chemical compositions to the values of the model before the CE phase does not change our main results for the $\alpha$- and $\gamma$-values. We plan to look into this issue in detail later, because it may be important for the seismic behavior of these stars.

We have determined the minimum core mass for helium ignition independent from \citet{han2002}, because we use a different definition of the helium core mass. Moreover, we also are interested in how much of the hydrogen envelope is allowed to remain after CE ejection, so that we can make a consistent estimate of the binding energy of the ejected envelope. Our results are in general consistent with theirs, except for the  minimum core mass for the helium flash. The reason is that we strip off more hydrogen, to exclude models that still have a growing core after CE ejection, because the star is then too large to fit in a narrow post-CE orbit. Therefore, we find slightly higher minimum core masses for the helium flash than \citet{han2002}.

Further, we observed that the models that can ignite helium when the envelope is lost during the Herztsprung gap, i.e.~those with $M_{\rm{ZAMS}}>2.5$ M$_{\odot}$, tend to expand even when we strip off the hydrogen envelope down to $X=0.10$, because, at this stage, the hydrogen profile is rather smooth due to the shrinking convective core during the MS. So there is still some hydrogen left, mixed in the core, which continues to burn after CE ejection, causing the star to expand further. An interesting consequence is that the stars ejecting the CE at this earlier stage, might have more helium in their remaining envelope. We also plan to examine this possibility in detail in future work.

\section{Conclusions}\label{conclusions}
We have studied the stellar structure of the progenitors of sdB stars in short-period binaries. The narrow range a giant's core mass can have to still ignite helium after loss of its envelope, allows us to estimate the orbital separation at the onset of mass transfer. This enables us to constrain the CE phase. We have compared two different CE-descriptions; the $\alpha$-formalism based on the energy equation (implicitly assuming angular momentum conservation), and alternatively, the $\gamma$-formalism based on the angular momentum equation (implicitly assuming energy conservation). Although we focused our study on the interesting system {PG\,1336$-$018}, the methods we present here can readily be applied to other sdB stars  formed in the CE ejection channel. In particular, Table \ref{mincore} is very useful for such studies.

Adopting the $\alpha$-formalism implies that  {PG\,1336$-$018} ignited helium in a degenerate flash. The sdB mass must then be between $0.39$ and $0.48$ M$_{\odot}$. However, the results are less convincing if the internal energy contributed to the ejection of the envelope. Furthermore, the $\gamma$-formalism does not rule out the possibility that the sdB progenitor ignited helium under non-degenerate conditions. In these cases, the possible mass range of the sdB star is wider: $0.3$ - $0.8$ M$_{\odot}$. If the CE is caused by a tidal instability, the $\alpha$-formalism might be preferred. If the CE is formed by runaway mass transfer, the $\gamma$-formalism is perhaps more likely \citep{nelemans2000}. Unfortunately, a detailed physical description for the CE evolution is still missing and these conditions are not definite. For the time being, we consider both formalisms to be valid.

We plan to use the pulsational properties of  {PG\,1336$-$018} to make an independent high precision mass determination of the sdB star and its envelope. Clearly, this will shed more light on the poorly understood CE phase. Also, if we are able to determine whether the sdB experienced a helium flash or ignited helium quiescently, this would provide us with much insight into the CE evolution. It is not clear however, if the helium flash leaves a (seismic) detectable imprint on the sdB interior. We will investigate these questions in a follow-up paper.  

\begin{acknowledgements}
We are indebted to P.~Eggleton for his scientific generosity. We also thank E.~Glebbeek and R.~Izzard for sharing their insight into the Eggleton code and WTTS. We are grateful to the referee for his/her valuable comments.
HH acknowledges a PhD scholarship through the ``Convenant Katholieke Universiteit Leuven, Belgium -- Radboud Universiteit Nijmegen, the 
Netherlands''. CA acknowledges financial support from the ``Stichting 
Nijmeegs UniversiteitsFonds (SNUF)'' and the Netherlands Research School for Astronomy (NOVA). HH, R\O, CA and MV are supported by the Research Council of Leuven University, through grant GOA/2003/04. GN is supported by NWO-VENI grant 639.041.405 and PJG by NWO-VIDI grant 639.042.201.
\end{acknowledgements}

\bibliographystyle{aa}
\bibliography{7133}
\end{document}